# Microsoft TerraServer: A Spatial Data Warehouse

Tom Barclay
Jim Gray
Don Slutz

Microsoft Research
Advanced Technology Division
Microsoft Corporation
One Microsoft Way
Redmond, WA 98052

# Microsoft TerraServer: A Spatial Data Warehouse

Tom Barclay. Jim Gray, Don Slutz
{TBarclay, Gray, Dslutz}@microsoft.com

Microsoft Research, 301 Howard St., Suite 830,San Francisco, CA 94105
http://research.microsoft.com/barc

## Abstract

Microsoft® TerraServer stores aerial, satellite, and topographic images of the earth in a SQL database available via the Internet. It is the world's largest online atlas, combining five terabytes of image data from the United States Geological Survey (USGS) and SPIN-2. Internet browsers provide intuitive spatial and text interfaces to the data. Users need no special hardware, software, or knowledge to locate and browse imagery. This paper describes how terabytes of "Internet unfriendly" geo-spatial images were scrubbed and edited into hundreds of millions of "Internet friendly" image tiles and loaded into a SQL data warehouse. Microsoft TerraServer demonstrates that general-purpose relational database technology can manage large scale image repositories, and shows that web browsers can be a good geospatial image presentation system.

## 1. Overview

The TerraServer is the world's largest public repository of high-resolution aerial, satellite, and topographic data. It is designed to be accessed by thousands of simultaneous users using Internet protocols via standard web browsers.

The TerraServer is a multi-media data warehouse. It differs from a traditional data warehouse in several ways: (1) it is accessed by millions of users, (2) the users extract relatively few records (thousands) in a particular session and, (3) the records are relatively large (10 kilobytes). By contrast, classic data warehouses are (1) accessed by a few hundred users via proprietary interfaces, (2) queries examine millions of records, to discover trends or anomalies, (3) the records themselves are generally less than a kilobyte. In addition, classic data warehouse queries may run for days before delivering results. Initial results typically cause users to modify and re-run queries to further refine results.

One thing the TerraServer has in common with classic data warehouses is that both manage huge databases: several terabytes of data. Terraserver's topographic maps cover all of the United States at 2 meter resolution 10 million square kilometers), the aerial photos cover 30% of the United States today (3 million square kilometers at one-meter resolution, and 1% of the urban areas outside the United States (1 million square kilometers) at 1.56 meter resolution.

This report describes the design of the TerraServer and its operation over the last year. It also summarizes what we have learned from building and operating the TerraServer.

Our research group explores scaleable servers. We wanted first-hand experience building and operating a large Internet server with a large database and heavy web traffic. To generate the traffic we needed to build an application that would be interesting to millions of web users. To have a huge database, we needed a huge data source: trillions of bytes that are relatively inexpensive to acquire and process.

Based on our exposure to the EOS/DIS project [Davis94], we settled on building a web site that serves aerial, satellite, and topographic imagery. We picked this application for three reasons:

1. The web is inherently a graphical environment, and these images of neighborhoods are recognizable and interesting throughout the world. We believed this application would generate the billions of web hits needed to test our scaleabilty ideas.
2. The data was available. The USGS was cooperative, and since the cold war had ended, other agencies were more able to share satellite image data. The thaw relaxed regulations that had previously limited the access to high-resolution imagery on a global basis.
3. The solution as we defined it – a wide-area, client/server imagery database application stored in a commercially available SQL database system – had not been attempted before. Indeed, many people felt it was impossible without using an object-oriented or object-relational system.

This paper describes the application design, database design, hardware architecture, and operational experience of the TerraServer. The TerraServer has been operating for a year now. We are just deploying the third redesign of the database, user interface, and online image loading system.

# 2. Application Design

Microsoft TerraServer is accessed via the Internet through any graphical web browser. Users can zoom and pan across a mosaic of tiles within a TerraServer scene. The user interface is designed to function adequately over low-speed (28.8kbps) connections. Any modern PC, MAC, or UNIX workstation can access the TerraServer in this way. If you have never used it, look at the TerraServer web site at http://terraserver.microsoft.com/.

Imagery is categorized into "themes" by data source, projection system, and image type. Currently, there are three data themes:

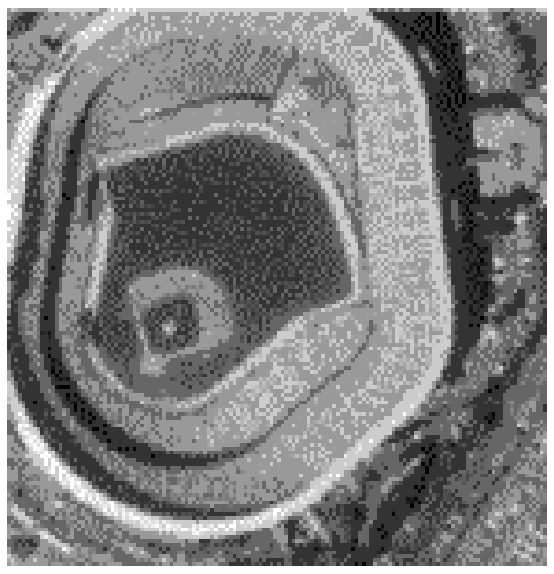

Figure 1. A USGS DOQ Image of 3Com Park near San Francisco

**USGS Digital Ortho-Quadrangles (DOQ)** are gray-scale, 1-meter resolution aerial photos. Cars can be seen, but 1-meter resolution is too coarse to show people. Imagery is orthorectified to 1-meter square pixels. Approximately 50% of the U.S. has been digitized. The entire conterminous U.S. is expected to be completed by the end of 2001.

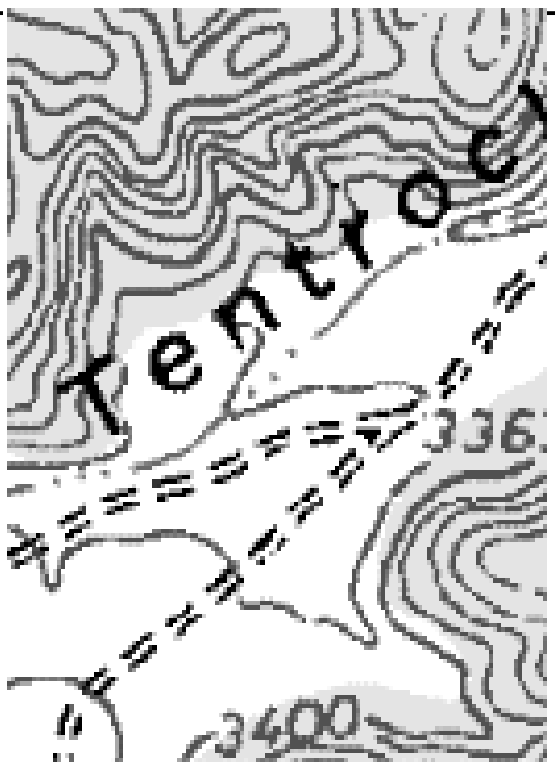

Figure 2. A USGS DRG 2-

**USGS Digital Raster Graphics (DRG)** are 13-color digitized topographic maps, with scales varying from 2.4 meter resolution to 25.6 meter resolution. DRGs are the digitized versions of the popular USGS topographic maps. The complete set of USGS topographic maps have been scanned including Alaska, Hawaii, and several territories such as Guam and Puerto Rico.

**Aerial Images SPIN-2™** are grayscale 1.56 meter resolution Russian satellite images. The images are re-sampled to 1-meter resolution. Microsoft TerraServer contains SPIN-2 images of Western Europe, the United States, and the Far East. Unfortunately, there is little coverage of Canada, South America, Africa, and Southeast Asia.

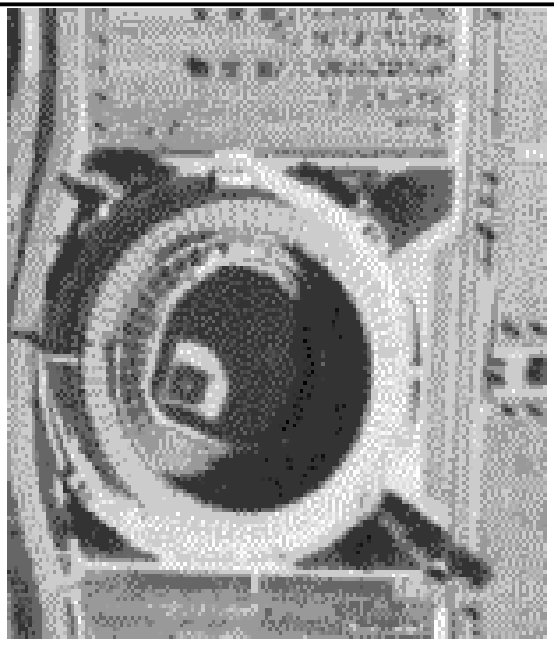

Figure 3. a SPIN-2: 1.56-meter image of Atlanta's Fulton County Stadium.

## *2.1 Projection Systems and Scenes*

The earth is a bumpy ellipsoid. Maps and computer monitors are flat. It is impossible to accurately present a spherical object on a flat surface.

Cartographers have addressed this issue by developing projections of the geoid onto flat surfaces [Robinson95]. There are many projection systems, each suited to present certain regions or properties. Multiple images in a projection system can often be joined together to form a seamless mosaic within certain boundary conditions. These mosaics either have extreme distortion as they scale out, or they introduce seams.

DOQ and DRG data are projected by the USGS into Universal Transverse Mercator (UTM) projection using the North American Datum (NAD) ellipsoid created in 1983 [Snyder89]. UTM is a projection system that divides the earth into 60 wedge shaped *zones* numbered 1 thru 60 beginning at the International Date Line. Each zone is 6 degrees wide and goes from the equator to the poles. UTM grid coordinates are specified as zone number, then meters from the equator and from the zone meridian[1].

The conterminous United States is divided into 10 zones (see Figure 4). Each of these UTM zones is a *scene*. The TerraServer mosaics each scene, but two adjacent scenes are not mosaiced together. Users can pan and zoom within a scene, and can jump from one scene to another.

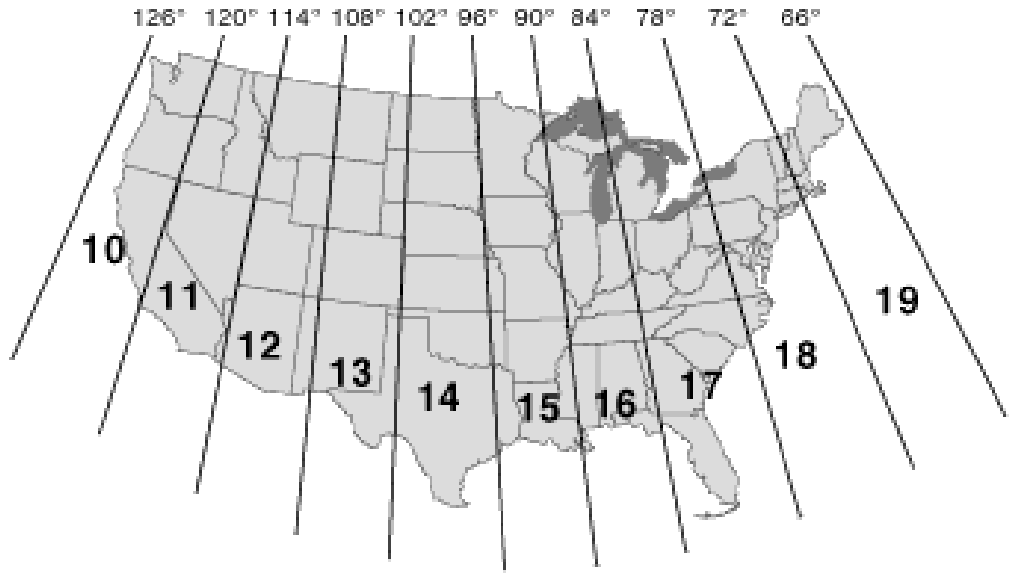

Figure 4: The ten UTM zones in the continental United States.

The Russian SPIN-2 imagery is digitized from Russian satellite photographs. The Russian satellite captures 160km wide by 40km high areas in a single image. The satellite takes one image and then begins the adjacent image, overlapping the last image. The overlap is variable, and when digitized does not line up on a pixel boundary.

To create a seamless mosaic of SPIN-2 imagery, all SPIN-2 imagery would have to be orthorectified. This requires precise geo-location of each image, which was not possible due to security concerns. Without rectification, if tiles extracted from separate SPIN-2 satellite images are mosaiced, the tile edges are misaligned. Roads, rivers, and other geographic features do not line up. While GIS experts may tolerate this, it is disorienting and unacceptable to novice users.

[1] Actually, UTM grid units can be in inches, feet, meters, or kilometers. The USGS chose meters for most of their assets in the UTM projection. UTM is not used above 80N or 70S [Robinson95].

Consequently, the TerraServer treats each 160km x 40km SPIN2 image as a separate scene. These scenes are not mosaiced together. Users can pan and zoom within a scene, and can jump from one scene to another.

## 2.2. TerraServer Grid System

Users can zoom and pan across a mosaic of tiles within a TerraServer scene. The tiles are organized in the database by theme, resolution, scene, and location within a scene.

TerraServer is designed to support a fixed number of resolutions in powers of 2 from 1/1024 meters per pixel (scale 0) through 4096 meter (scale 22). The scale is related to resolution in meters per pixel by

$$\text{Scale} = \log_2(\text{resolution}) + 10$$

The highest resolution images currently in the database are one meter per pixel, which is scale 10. Coarser resolutions are derived by sub-sampling fine-resolution images.

For UTM projection data-sets, the SceneID is the UTM zone assigned to the original image a tile's pixels were extracted from. For SPIN2 data-sets, a unique SceneID is assigned by TerraServer for as each scene is loaded.

Each TerraServer scene is planar. A tile can be identified by its position in the scene. The tile loading program assigns a relative X and Y tile identifier to each tile as it is loaded.

For UTM projected data, the X and Y tile address is the UTM coordinate of the top-left pixel in the tile divided by the tile image size in UTM units in meters. The following are the formulas:

X = TopLleftUTM_X / (TilePixWidth • Resolution)

Y = TopLeftUTM_Y / (TilePixHeight • Resolution)

For SPIN2 scenes, the X and Y tile addresses are relative to the upper left corner of the scene.

The six fields – Resolution, Theme, SceneID, X, and, Y - form the unique key by which any TerraServer image tile can be directly addressed. Each TerraServer web page contains image tiles from a single Theme, Scale, and SceneID combination. For example, our building in USGS DOQ theme (T=1), has scene UTM zone 10 (S=10), at scale 1 meter (Z=10) with X=2766 and Y=20913. The URL is: http://terraserverv.microsoft.com/tile.asp?S=10&T=1&Z=10&X=2766&Y=20913.

The TerraServer search system performs the conversion from geographic coordinate systems to the TerraServer coordinate system. The TerraServer image display system uses TerraServer grid system coordinates to pan and zoom between tiles and resolutions of the same theme and scene.

## 2.3. Imagery Database Schema

Each theme has an ***OriginalMeta*** table. This table has a row for each image that is tiled and loaded into the TerraServer database. The *OrigMetaTag* field is the primary key. The meta-fields vary widely from theme to theme. Some of the meta fields are displayed by the Image Info Active Server Page (for example, see http://terraserver.microsoft.com/GetOrigMeta.asp?OrigMetaId=104578&SrcId=1&Width=225&Height=150&ImgSize=0&DSize=0)

All the image tiles and their metadata are stored in an SQL database. A separate table is maintained for each (theme, resolution) pair so that tiles are clustered together for better locality. USGS DOQs have resolutions from 1-meter through 64-meter. USGS DRG data supports 2-meter through 128-meter resolution. SPIN supports resolutions from 1-meter to 64-meter.

Each theme table has the same five-part primary key:

- *SceneID* –individual scene identifier
- *X* – tile's relative position on the X-axis
- *Y* – tile's relative position on the Y-axis
- *DisplayStatus* – Controls display of an image tile
- *OrigMetaTag* – image the tile was extracted from

There are 28 other fields that describe the geo-spatial coordinates for the image and other properties. One field is a large "blob" type that contains the compressed image.

These tile blobs are chosen to be about ten kilobytes so that they can be quickly downloaded via a standard modem (within three seconds via a 28.8 modem).

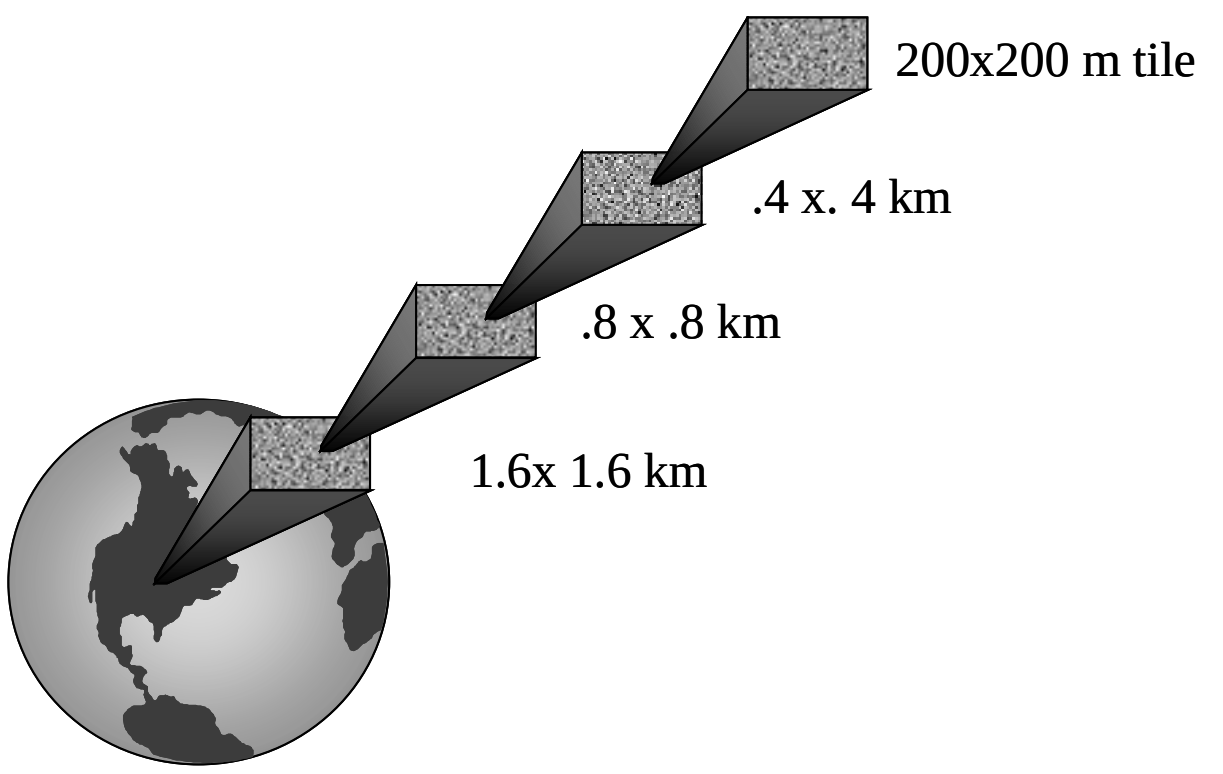

Figure 5: The image database is organized as a pyramid of 200x200 pixel tiles. Coarser images are sub-sampled from fine-resolution images.

## 2.4. Gazetteer Database Schema

The Gazetteer lets users find images by name. It contains the names for about 1.5 million places, with many alternate spellings. It is a simplified version of the Gazetteer found in the Encarta Virtual Globe™ and Microsoft Streets™ products.

The Gazetteer Schema has a snowflake structure. *Place* is the center table. It contains the formal name for a unique place on earth and maps the uniquely named location to the TerraServer Grid System. The *AltPlace* table contains all the synonyms of a unique place. The *State* and *Country* parent tables identify a place's state/province and country. The *AltState* and *AltCountry* tables contain the state/province and country synonyms.

Lookup by place name is surprisingly common (40%). So the user interface was modified to make it even easier. The TerraServer home page has a simple name lookup field, and a button that takes the user to an "advanced" name lookup web page. The *find a place* input field allows the user to enter a subset of *place name*, *state name*, and *country name*. The supporting database stored procedure builds a SQL cursor that searches for the name by performing joins on the appropriate tables, depending on which fields the user specified. Name searches are performed on the "Alt" tables which have synonyms and abbreviations for places (USA for example). Formal names matching the search criteria are returned from the *Place, State,* and *Country* tables.

The *ImageSearch* table forms the association between a named place and visible images. The *ImageSearch* table identifies the Theme, *SceneID, Scale, X, Y*, and *ImageDate* of visible image tiles that cover the kilometer-square cell at the center of the named place. The load program inserts rows into the *ImageSearch* table when the sub-sampling program completes filling in the image pyramid for a certain area. The *ImageSearch* table serves as a one-level quad-tree index of the image data [Samet90].

The image display Active Server Page scripts use an additional table, the *Pyramid* table, to display the city closest to the center tile on an image display web page. This two-level quad-tree is used to find population-weighted nearest neighbors of a given latitude and longitude. The SQL stored procedure scans a rectangle of the quad-tree to determine the closest city. The quad tree is implemented atop a B-tree by giving each quadrangle a name which is a prefix of the key for records in that quadrangle.

In total, the Gazetteer contains about 4 million rows and consumes 3.3 GB of space. Our first design used a fine-granularity (quarter kilometer) quad-tree and so used a hundred times more space 400GB). That design needed no computation during lookup. The current design must examine 50 records on average and evaluate some spherical trigonometry on the coordinates for each record. The new design uses more computation, but it can examine a record in 30 microseconds of processor time, so it is a good tradeoff.

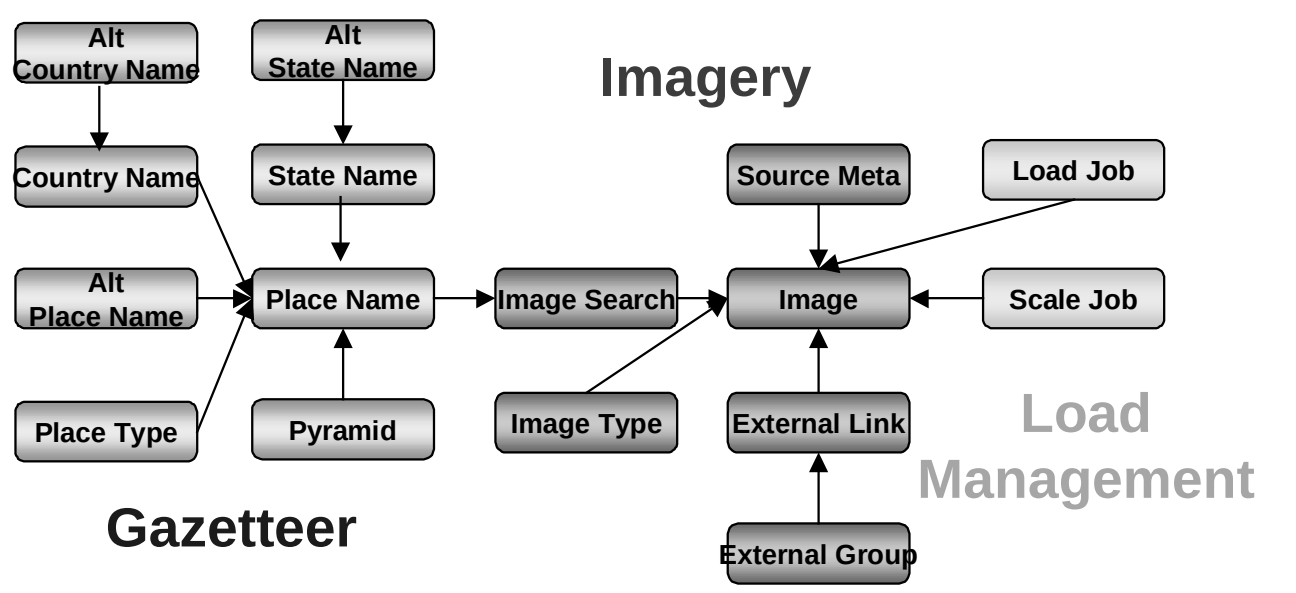

Figure 6. The gazetteer forms a star schmea used to locate places by name. The Pyramid is a quad-tree used to located the nearest city. Most of the data is stored in the Image database which is indexed by theme, resoulution, sceneID, scale, X, and Y. Source Meta represents the OriginalMeta table. The job tables track the data loading process and the data sub-sampling to build the image pyramid.

# 3. TerraServer User Interface

The TerraServer user interface is designed to be useable by a sixth grade Geography student. We wanted users to naturally understand how to find and view images with a minimum of instruction. As with video games – practically everything in the image viewing area is click-able.

The image location methods all display the result as a web page formatted with a table of tiles at the middle of the image pyramid. The user can zoom-in to move down the image pyramid or pan in any direction – Northwest, North, Northeast, West, East, Southwest, South, and Southeast. Users can also zoom out from lower levels in the image pyramid.

The image display web page is a simple HTML document containing a table of image source tags identifying the specific image tiles to form the picture. Anchor tags allow the user to pan and zoom through the tiles of a single theme at a time. Any web browser that supports HTML tables and can display Jpeg and Gif images can host the TerraServer user interface. Full resolution SPIN-2 imagery requires the web browser to support Java applets.

Automated processes can also access the imagery via the HTTP protocol. This enables TerraServer imagery to be integrated into third party applications. The USGS and SPIN-2 data providers use this feature to sell and distribute their imagery on-line. TerraServer routes image "download" requests to data provider web sites. The data provider commerce applications fetch Microsoft TerraServer image tiles programmatically during the image purchase and image delivery process.

## 3.1. Navigation

Users have several methods to navigate to an initial spot on the globe. From there they can pan and zoom within a scene, or jump to a new scene.

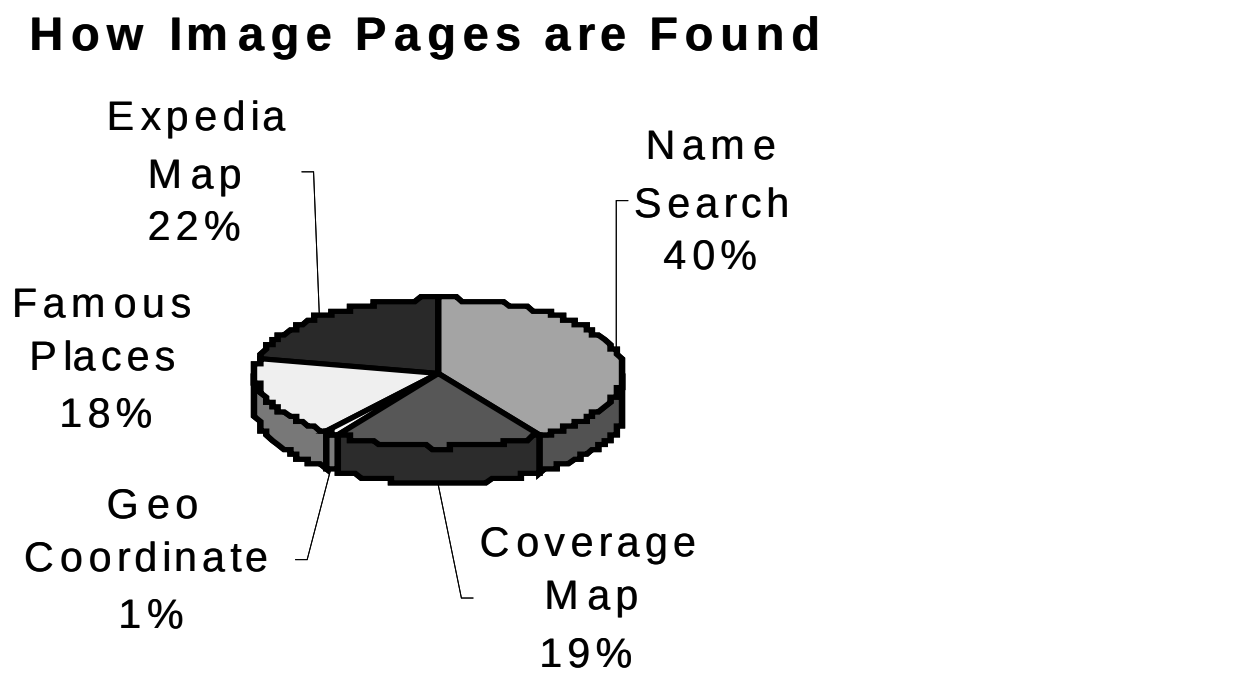

Figure 7: Name search is the most popular way of finding a place.

**Name Lookup:** Clients knowing the place name (or part of the name) can navigate textually by presenting a name to the Gazetteer. The Gazetteer database has the names and locations of 1.5 million places in the world. For example, *Moscow* finds 28 cities, while *North Pole* finds 5 cities, a mining district, a lake, and a point-of-interest. There are 378 entries for San Francisco in the Gazetteer. All entries with matching images are shown in a list along with the data provider name and the image date. The user selects an image from the list.

**Expedia maps:** To help orient users, a political and street map can be displayed in place of the image. Users can navigate by clicking on the map to center and zoom-in or pan in any direction. Once the user centers the map over their point of interest, he can switch back to the imagery.

**Famous Place List:** The Famous Place list is a set of direct links to familiar places (like the Pyramids). Users suggest entries for the Famous Places List. TerraServer administrators review the suggested places and add entries to the Famous Place table in the TerraServer database. Thumbnail images of some of them are added to the TerraServer home page for easy access. Clients interested in viewing well known imagery of major cities, natural wonders, and famous sites can browse a pre-selected list of links to them.

**Coverage map:** The USGS provided us with a set of base maps in three resolutions – 1 pixel per degree, 8 pixels per degree, and 48 pixels per degree. The TerraServer database load programs color in areas of these three *coverage maps* as data is loaded into the system. The result is a shaded map depicting where there is image coverage. Three separate coverage maps are produced: all data providers, SPIN-2 only, and USGS only. The coverage map system is useful to those seeking an image “near New York City”, assuming users have an idea where New York City is without needing to name it.

**Navigation via latitude and longitude:** Some users have a GPS or other source of precise geographic reference information. The TerraServer supports direct entry of latitude and longitude values in this case.

As Figure 7 shows, most lookups are by place name. Map-based navigation is the second-most popular method. The famous-place list and coverage map are tied for third place.

## 3.2 TerraServer Image Page

TerraServer search operations identify the center tile of the image display page. The image display system builds a complete web page of image source and anchor tags around this tile. In addition to the mosaic of tiles, the page includes information aids that explain what the user is viewing clickable buttons for navigation (see Figure 8):

- The relative distance from nearest city, e.g. 20 Km SW of Raleigh, North Carolina, United States.
- The image date.
- A list of other images overlapping the current image.
- The logo of the data source, e.g. USGS or SPIN-2.
- The resolution, e.g. 64m, 32m, 16m, 8m, 4m, 2m, 1m etc.
- *Zoom In/Out* buttons to move in or out one resolution level from the current level. Clicking directly on an image zooms in one resolution level and pans the clicked image to the center of the image display area.
- *Pan* buttons (bright green triangles) around the border of the image window. The button is grayed out if imagery is not available in a direction.
- Three buttons control the size of the image display area – Small, Medium, and Large. Small is designed to fit an 800 x 600 resolution monitor. Medium is designed to fit a 1024 x 768 monitor. And Large is designed to fix 1280 x 1024 and larger resolution monitors.
- The *View Map* button switches the display to a Microsoft Expedia street map centered over the image area.
- The *Download* button links to a data provider web site where the user can purchase a digital or print copy of the image.
- The *Image Info* button links to a page describing the attributes of the source image the tiles on the web page were extracted from.
- The scale of the image, e.g. an English Mile/Yard scale and a Metric Kilometer/Meter scale.

The user interface has evolved over the last 2 years, and continues to evolve. It is likely that it will have changed by the time you read this.

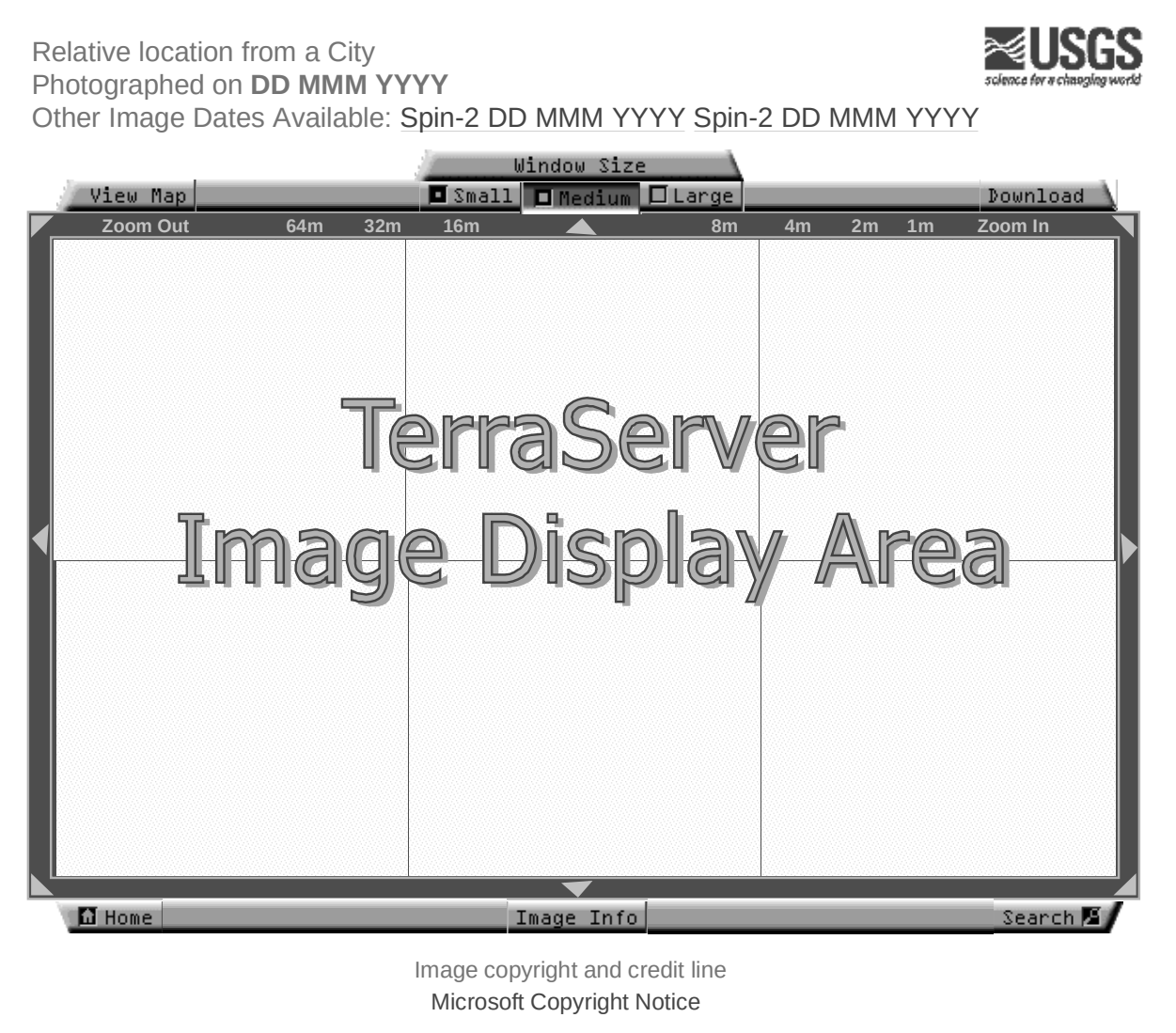

Figure 8 The TerraServer image frame. The image display area contains one or more tiles (default is 2x3).

# 4. System Architecture

## 4.1. Three-Tier Software Architecture

The Microsoft TerraServer has a 3-tier architecture. As depicted in Figure 8, the darker shaded boxes identify standard “off-the-shelf” software. The lighter shaded boxes identify application software that implements the TerraServer application logic.

**Tier 1:** *The Client* is a graphical web browser or other hardware/software system that supports HTTP 1.1 protocols and HTML 3.2 document structure. Microsoft TerraServer is built and tested with Netscape Navigator V3.0 & V4.0 and Internet Explorer V3.0, V4.0, & V5.0 on Windows, MacOS, and UNIX.

**Tier 2:** *The Application Logic* is a web server application that responds to HTTP requests submitted by clients by interacting with the Tier 3 database system and applying logic to the results returned.

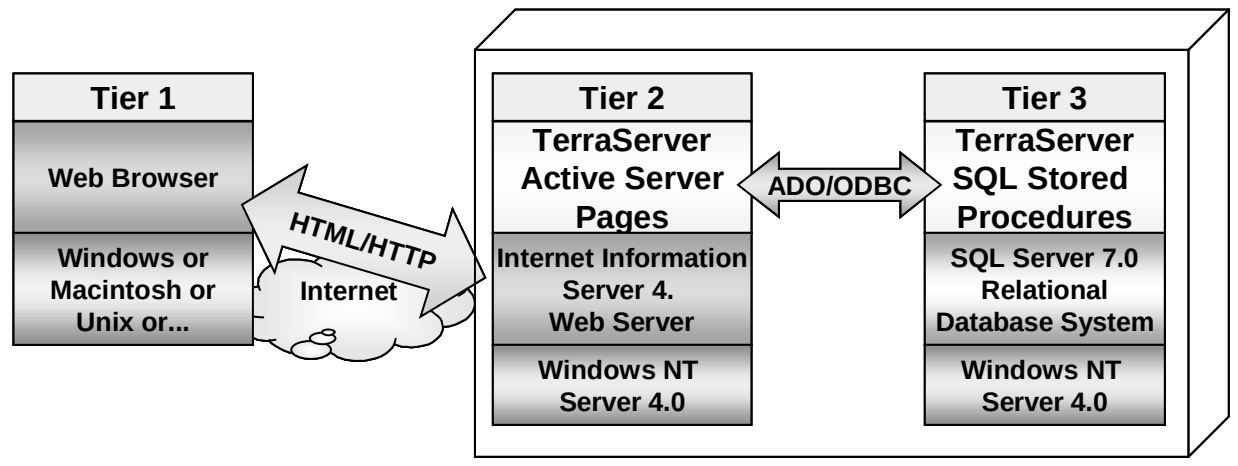

Figure 9: The TerraServer has a 3-tier architecture.

**Tier 3:** *The Database System* is a SQL Server 7.0 Relational DBMS containing all image and meta-data required by the Application Logic tier.

All end user access to TerraServer image and metadata occurs through Tier 2 stored procedures (Active Server Pages[2]) written in Visual Basic Script (VBScript). Thirty two Active Server Page (ASP) scripts implement the entire TerraServer web application. These ASP scripts dynamically construct HTML documents from information returned by SQL Server stored procedures. The scripts invoke SQL Server stored procedures via an Active Data Object (ADO) interface that is layered on top of the Open Data Base Connection (ODBC) protocol.

ASP scripts return complete HTML documents to web browser clients. Each web browser decodes the HTML and formats the user’s browser window. The web browser opens the image tile URL that causes the web server to invoke the image fetch ASP script.

All meta and image data is stored in a single Microsoft SQL Server 7.0 database. Multiple database servers can be configured into a single Microsoft TerraServer web site. The only rule is that all data for a theme must be contained entirely in one database. A theme’s data can be replicated to one or more backup servers, but with our current stored procedures, a theme’s inventory cannot be split across two database servers.

The TerraServer SQL Stored Procedures perform the application’s data access logic so that each application function has only one round-trip from the web page to the database server. There are twelve stored procedures to support online access and eight stored procedures to support the database loading process.

## 4.2. Database Architecture

The database architecture was chosen to demonstrate the scalability and usability of SQL Server—everything was done in the most straightforward way, with no special tricks A single SQL server database was created on one file group consisting of many NTFS files. Each file resided on one of the four logical volumes and was 20GB which is a convenient backup/recovery unit. Initially, 53 files were created to achieve the 1TB database goal. Additional files are added as new imagery is loaded. Plans are to grow the database to over 2.2TB. The initial files were placed on two 595GB NT stripe-set volumes and the files added later were placed on two other similar volumes. SQL Server makes all allocation and placement decisions within and among the files.

The TerraServer database was created using default settings with two exceptions. A bulk copy option was set to improve load times by reducing logging. Also, a truncate log on checkpoint option was set. These options preclude media recovery using the log. Instead, Terraserver would restore from an online database backup and reload any data that had been added since that backup.

The physical SQL table design closely follows the logical schema of Figure 6. There is a separate set of Image tables for each *theme*. The set of tables is almost identical for each *theme* and consists of an ***OriginalMeta*** table with attributes of the *theme* for each original image and a set of ***Image*** tables, one for each scale. An Image table contains a row for each image tile of that scale and the primary key is the Scene identifier, X-axis coordinate, Y-axis coordinate, display status, and OriginalMeta primary key value. The tile image, about 10KB average size, is placed directly in a column of type *image*, there is no image data outside the database. The tile image size was set solely by the application needs and just happens to be a little larger than the SQL Server page size of 8KB. SQL Server automatically manages the mapping of image column data onto pages. The image column comprises almost all the bytes in the row and the Image tables comprise almost the entire database contents.

All tables are clustered on their primary key and a few secondary indexes, mostly in the Gazetteer, were added to support searching for different name combinations and for on-line loading. Retrieving one image tile requires the simplest of SQL statements:

```
Select * from Image where PrimaryKey='value'
```

One set of Gazetteer tables and one *ImageSearch* table serve to locate images by name in all themes. The *Loadjob* and *Scalejob* tables are used to manage the online loading of images. They hold the state of load jobs and are used for monitoring and restart.

[2] Active Server Page is technology similar to CGI scripts in other web server implementations. Active Server Page scripts have better performance characteristics than CGI.

## 4.3. Commerce Server

The data source providers – USGS and SPIN-2, own the imagery stored within the Microsoft TerraServer database. The USGS and SPIN-2 organizations, not Microsoft, offer a private use license of their imagery for a nominal fee. Users purchasing USGS data receive one or more “Digital Ortho Quarter Quadrangle” data files on a single CDROM through the US Mail. Users purchasing SPIN-2 data receive a digital file through the Internet and can optionally purchase a photographic print from Kodak.[3] The USGS and SPIN-2 each host their own electronic commerce web sites.

The USGS commerce web site is located the USGS EROS Data Center in Sioux Falls, South Dakota. The web site runs Microsoft Site Server Enterprise Edition 3.0 on Intel based equipment. Custom components connect the MS Commerce application with internal USGS accounting and CD fulfillment systems. USGS EROS Data Center personnel developed the USGS commerce application and the components that interconnect the commerce server with internal applications.

The SPIN-2 commerce web site is located at Aerial Images’ headquarters in Raleigh, North Carolina. The web site runs Microsoft Site Server V3.0 on a two-node Compaq Alpha-Cluster. Microsoft developed the commerce application and the interconnection to Kodak for the rights to display SPIN-2 data on the Internet.

The TerraServer web site hosts a store-front to the USGS commerce web site. The first web page explains the difference between an “Internet ready image file” like a Jpeg and the full-resolution USGS DOQ file. The user is offered a free, Internet-ready, Jpeg file in case the user does not have the software or interest in purchasing the more technically challenging USGS DOQ data file. If the user selects a free download, then the TerraServer web site downloads the ImageSave java applet that creates a single Jpeg file of a USGS DOQ image on the user’s hard drive. The user is returned to the TerraServer web page where they clicked the “Download” button.

If the user decides to purchase a USGS DOQ, the TerraServer store-front lists the USGS DOQ files used to create the imagery on the web page where the user clicked the “Download” button. The user can click check-box buttons to reduce the number of DOQ files purchased. Clicking “Purchase” causes TerraServer to pass the list of requested DOQ files to the USGS commerce application running at the EROS Data Center. At the USGS commerce site, the USGS commerce application adds the DOQ files to the user’s shopping basket. The user can edit the shopping basket, proceed to the “checkout register”, or return to TerraServer.

## 4.4. Hardware Architecture

The web site is configured to minimize single points of failure, protect the database from hackers, and scale to support additional users or data over time.

[3] SPIN-2 has an arrangement with Kodak to produce a photographic print of SPIN-2 imagery purchased through TerraServer.

The Tier 2 and Tier 3 software runs on separate computer systems. There is an HTTP firewall in front of the web servers and a packet filter firewall between the web servers and the database server. Having the database server inside the corporate firewall allows us to load data to the TerraServer from within the Microsoft corporate network.

The web site has seven WindowsNT servers. The database system is a Compaq AlphaServer™ 8400 containing 8 440 MHz Alpha processors and 10 GB of RAM. The processor is attached to 7 StorageWorks™ Enterprise Storage Array 10000 (ESA-10000) cabinets. The disk arrays are based on UltraSCSI technology.

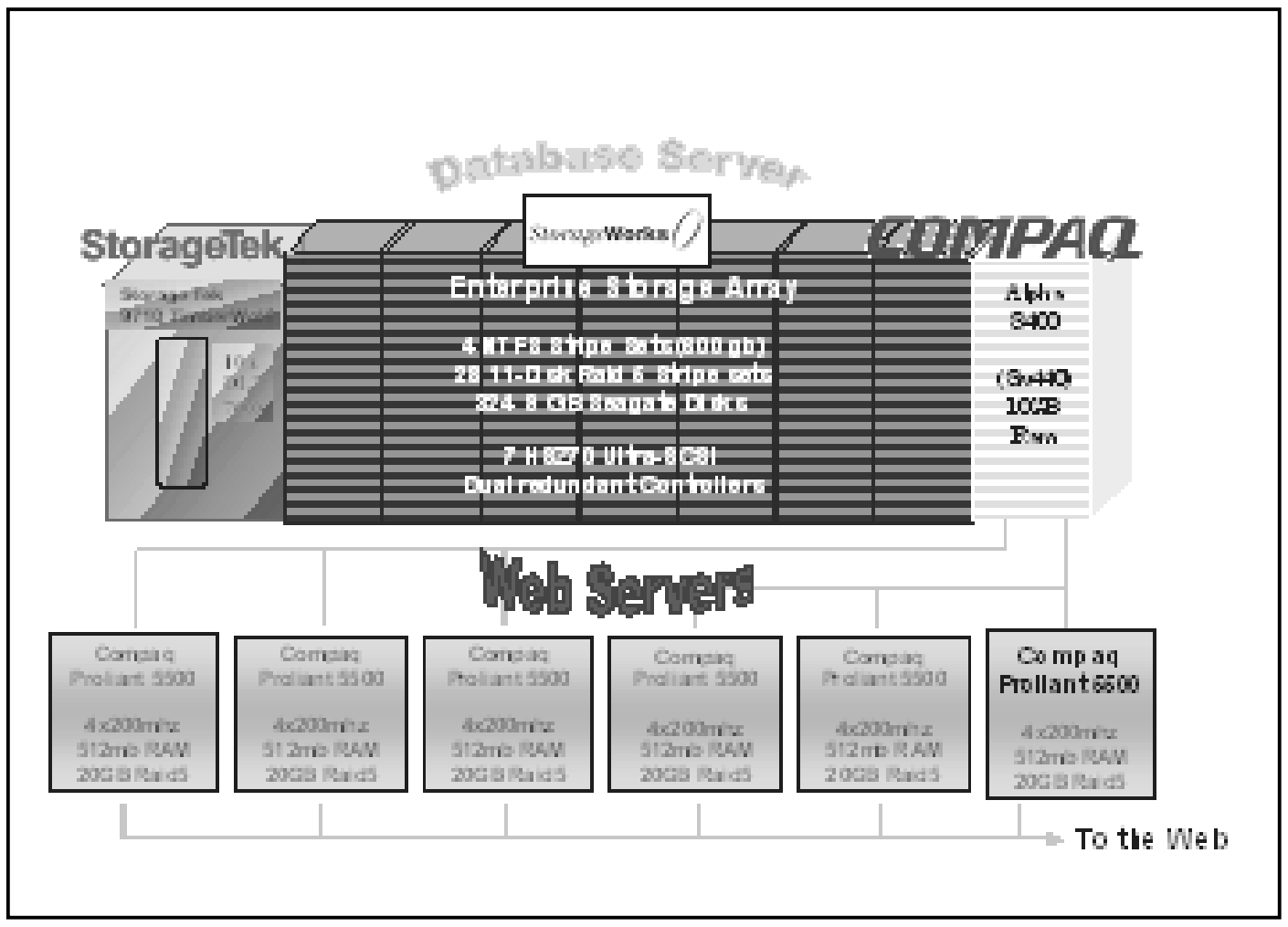

Each ESA-10000 contains 46 9 GB disk drives and 2 HSZ70 dual-redundant RAID-5 controllers. 4 sets of 11 disks each are configured into a single RAID-5 stripe-set and managed as a single logical disk by the HSZ70 controller. 2 drives per cabinet are available as hot spares. Should a disk fail, the HSZ70 controllers automatically swap a spare drive into a RAID set.

WindowsNT Server sees each 4 large disks (85 GB each) created by the RAID controllers of each of the seven disk cabinets. It stripes these into 4 large (595 GB) volumes which are then each formatted and managed by the WindowsNT file system (NTFS). Each 595 GB volume contains about thirty 20GB files. SQL Server stores its databases in these large files. We chose a 20GB file size since it fits conveniently on a magnetic tape.

Connected to the AlphaServer 8400 is a StorageTek 9710 automated tape robot. The tape robot contains 10 Quantum DLT7000 tape drives. Legato Networker backup software can backup the entire 1.1 TB TerraServer SQL database to the StorageTek tape robot in 7 hours and 15 minutes – or 17 GB/hour.

For reliability and performance, the database server contains three 100 Mbit Ethernet cards and is connected to three separate local area networks. One network card connects the database server to three of the Compaq ProLiant 5500 web servers. A second network card connects the database server to three other Compaq ProLiant 5500 web servers. The third network card connects the database server to the TerraServer image processing center which is inside the Microsoft corporate network.

The TerraServer web site is housed at the Microsoft Internet Data Center – a well-managed and secure facility with excellent environmental protection (emergency power, good physical security,...), and with high bandwidth to the Internet (about 8 Gbps at present).

### 4.5. Hardware Capacity Planning

It is difficult to size an Internet application in advance. We originally planned for three-million web hits per day, which is far beyond what we actually expected. At the time, other groups were reporting small numbers (e.g. 17 million hits per week for the 1997 winter Olympics.) But, publicity and interest in the site was very high. During the first week, demand was in excess of 30 million web hits per day. Ten times what we expected. This was not a pleasant experience for us or for our users.

Now that the novelty has worn off, demand averages 7 million hits per day with peaks of 15. The web site is configured to support a maximum of 6,000 simultaneous web browser connections and about 40 million web hits per day (see Table 1). Additional Tier 2 Web Servers could increase this number.

Table 1. TerraServer hardware configuration parameters

| | |
|---|---|
| Max hits per day | 40 million/day |
| Max SQL queries per day | 37 million/day |
| Max image downloads / day | 35 million/day |
| Bandwidth to Internet | 200 Mbps = 2 T Byte/day |
| Concurrent web connections | 6,000 connections |
| Web front ends | 6 4-way 200 MHz Compaq Proliant 5500, .5GB ram |
| Database back-end | 8-way 440MHz<br>Compaq AlphaServer 8400<br>10GB ram, 3.2 TB raid5<br>324 9GB Ultra SCSI disks |

# 5. TerraServer Data Load Process

As with other data warehouses, most of the labor of building the TerraServer consists of data scrubbing and data loading. The TerraServer database is organized to simplify the TerraServer web application that presents image and meta data to end users. The TerraServer design avoids dynamic projection, rotation, and other sophisticated features found in commercial GIS systems. The data loading programs precompute the GIS details and present each scene as a seamless mosaic of 200 pixel by 200 pixel tiles. All knowledge of projection systems, re-sampling pixels, edge alignment, merging pixels from multiple images, etc., is implemented in the load programs.

There are two image load programs in the TerraServer system – TerraCutter and TerraScale. *TerraCutter* re-formats imagery received from data sources, tiles it into formats acceptable to the TerraServer web application, and inserts the tiles and metadata into the database. *TerraScale* computes the lower resolution image-pyramid tiles for a theme by sub-sampling the tiles created by TerraCutter.

We implemented a simple job-scheduler system to manage and track the data loading process. Each processing program leaves a “popcorn trail” in the Load Management database so administrators can monitor progress on loading new data.

New imagery is inserted into the TerraServer database on-line while web users browse existing imagery. The table design and load program insertion order ensure that all the required metadata and imagery is in place before the image is made visible to the web application.

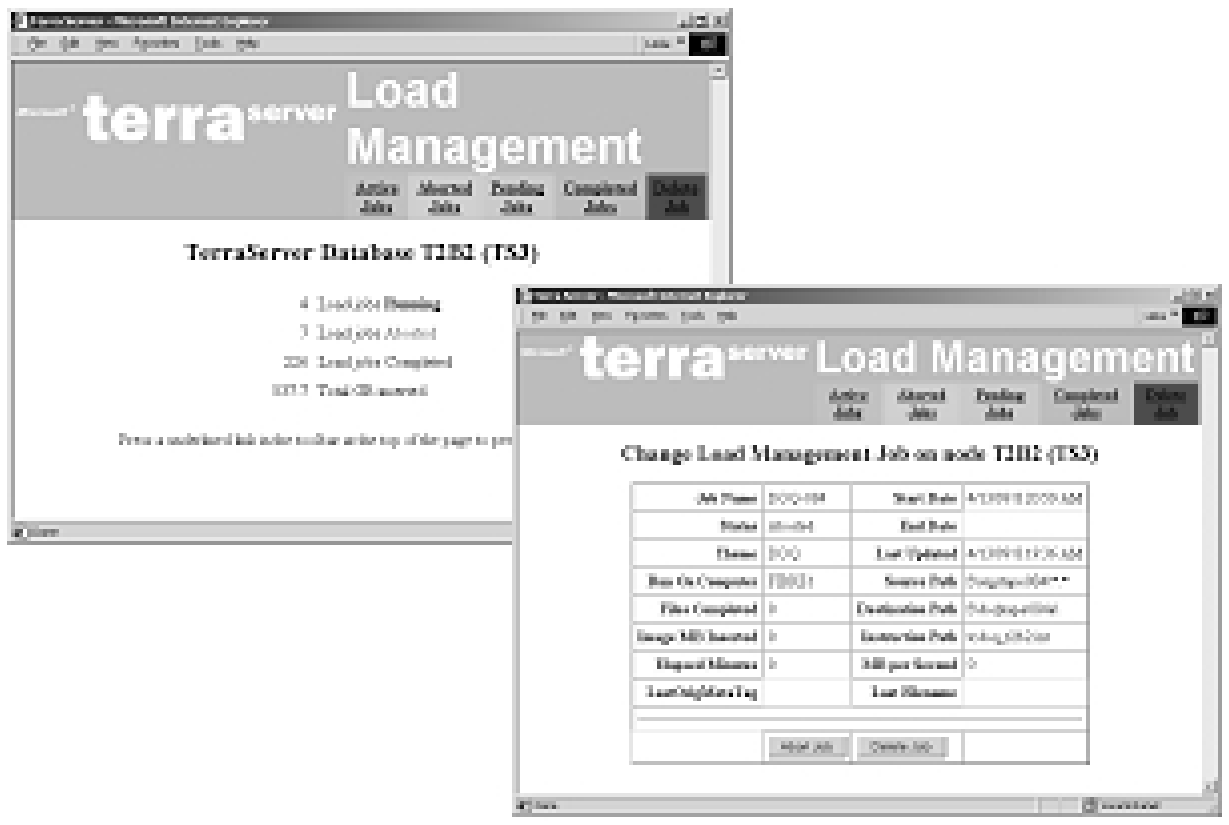

Figure 11. The load process is managed by two web ASP applications that start jobs and track their progress. The job status is recorded in the tables described in Figure 6 (and the text).

## *5.1. Data Flow*

USGS DOQ data is shipped to Microsoft via DLT media written in the “tar” format. DOQ files are in a custom USGS format. Meta-data and image pixels are contained in one file. Data is 8-bit grayscale or 24-bit, RGB color infra-red. TerraCutter converts color infra-red to 8-bit grayscale. DOQ files cover a USGS “standard quarter-quadrangle”, which is a 3.75 minute by 3.75 minute square area. The order of DOQ files on tape is random. Adjacent DOQ files can arrive in any order.

USGS DRG data is shipped to Microsoft on CDROM media. All 1:24,000, 1:100,000, and 1:250,000 scale maps for a square degree are contained on one CDROM. Images are in the GeoTiff format and have a fixed color map of 13 colors. Meta-data and image pixels are in separate files.

SPIN-2 data is shipped to Microsoft on DLT media written in “NT Backup” format. SPIN-2 files are in a custom “Kodak/Microsoft/Aerial Images” format. Meta-data and image pixels are in separate files. Data is 8-bit grayscale.

TerraServer System Administrators use the appropriate “off-the-shelf” program to download a tape or CDROM to a directory on one of six image editing systems.

Image editing systems are multi-processor Windows NT Server systems with 500 GB or more of local disk. Four servers are 4 processor 200 MHz Intel Servers donated by Intel. Two servers are 4 processor 300 MHz Alpha Servers donated by Compaq. Two Intel Servers are connected to 1 TB of Fiber-Channel disk array donated by Clariion, a subsidiary of Data General. The other two Intel servers are connected to two Symmetrix SCSI based disk arrays donated by EMC. The two Alpha servers are connected to a 250 GB StorageWorks disk array donated by Compaq. Each system has 4 to 6 100 GB stripe-set disk volumes.

The TerraServer System Administrators launch the TerraCutter image-editing program against a directory containing the image and meta files downloaded from tape or CDROM. TerraCutter uses the Load Management schema tables to make sure the job has not been processed previously. Or, if a previous run had aborted, TerraCutter will pick up where it had left off. TerraCutter uses the Load Management schema to catch duplicate files sent on previously processed tapes or CDROMs. When a directory has been successfully processed, the download directory is deleted, the tape is physically marked as “processed” and shelved. All further processing – sub-sampling to create lower resolution scales, correlating tiles with named locations, merging pixels between tiles, etc. – occurs within the memory of a custom program or T-SQL database statements.

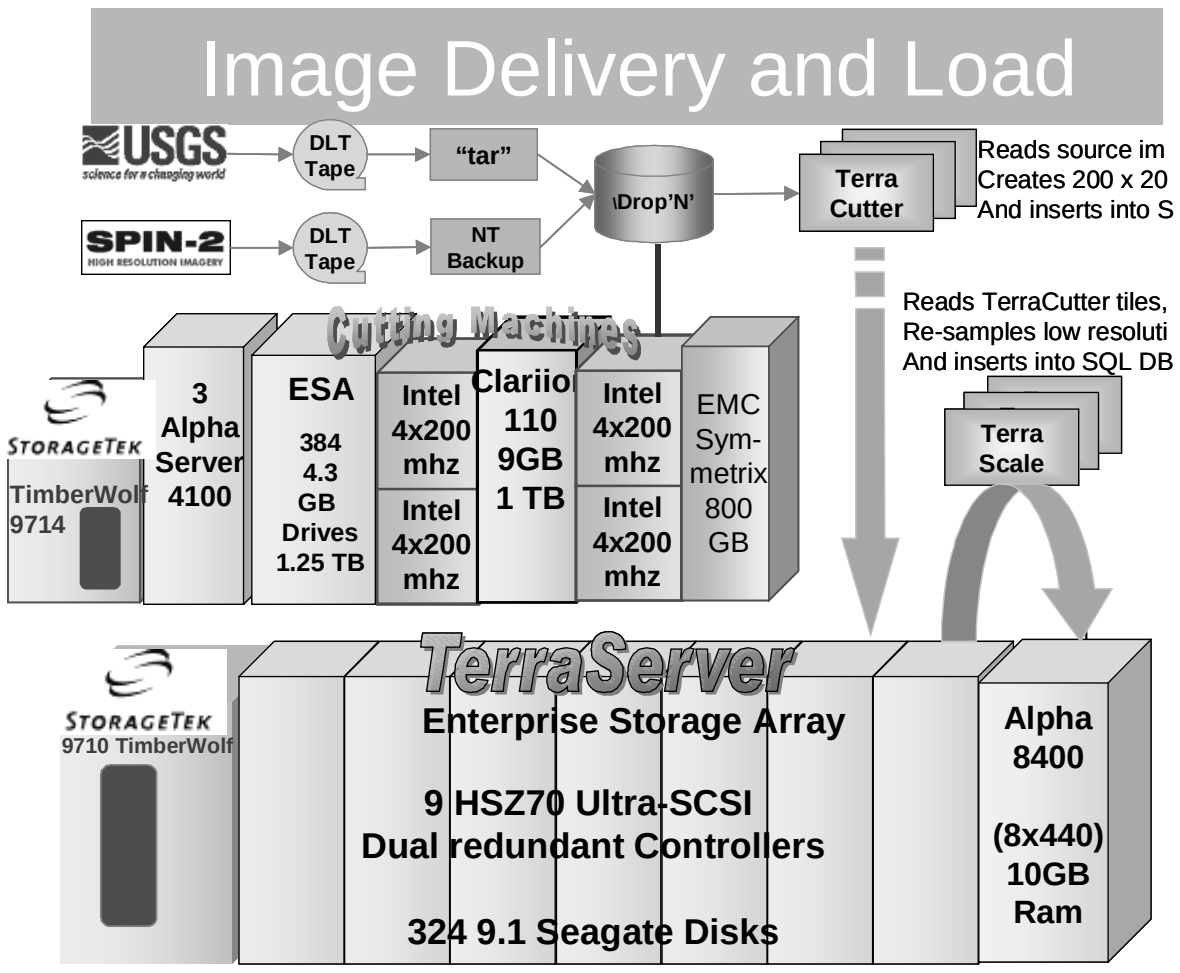

Figure 12. The image loading dataflow. TerraCutter populates the base of the image pyramid. Then TerraScale builds the image pyramid.

## 5.2. Load Management Schema

The Microsoft TerraServer load system maintains a set of tables in the TerraServer database. These tables are not visible to end users on the Internet. A set of Active Server Page scripts allow the TerraServer System Administrators to schedule and monitor the TerraServer database load process.

A *LoadJob* row is created when a load program is instructed to process a directory or a specific list of imagery received from a data source. The *LoadJob* row describes the on-disk location of the input data, the source tape/CD, the computer system the load program ran on, the load program version, the date the job started, and the job's current status.

Load programs update the *LoadJob* record each time they complete an input file found in the source path and insert a row into the *ScaleJob* table. This is the signal to the TerraScale program that a block of image tiles is ready to have its image pyramid created.

The TerraScale program updates the *ScaleJob* table with its progress information. There is a set of administrative Active Server Pages that TerraServer Administrators use to monitor the progress of image pyramid creation.

## 5.3. TerraCutter

TerraCutter is a fairly complicated C program. The simple part is formatting tiles suitable for the TerraServer web application and inserting them into the database. The TerraServer web application expects tiles to be in one of three formats:

- 8-bit Grayscale, Jpeg compressed
- 24-bit RGB, Jpeg compressed
- 2-8-bit, 4 to 256 color map, GIF compressed

The ground size covered by a pixel must also be fixed to multiples of 1-meter resolution –¼, ½, 1, 2, 4, 8, 16, etc. If necessary, TerraCutter re-samples the input image to the appropriate resolution as the image is read in. As tiles are produced, TerraCutter saves the tile image into a temporary file, computes the Image table metadata fields, and inserts the new tile into the database using ODBC API calls. A single image tile is inserted in the scope of one transaction.

The tiling process is the complicated part of the TerraCutter. Raw datasets are not mosaiced together, and so are easier to processes. Projected themes are more complicated and are handled differently than raw themes.

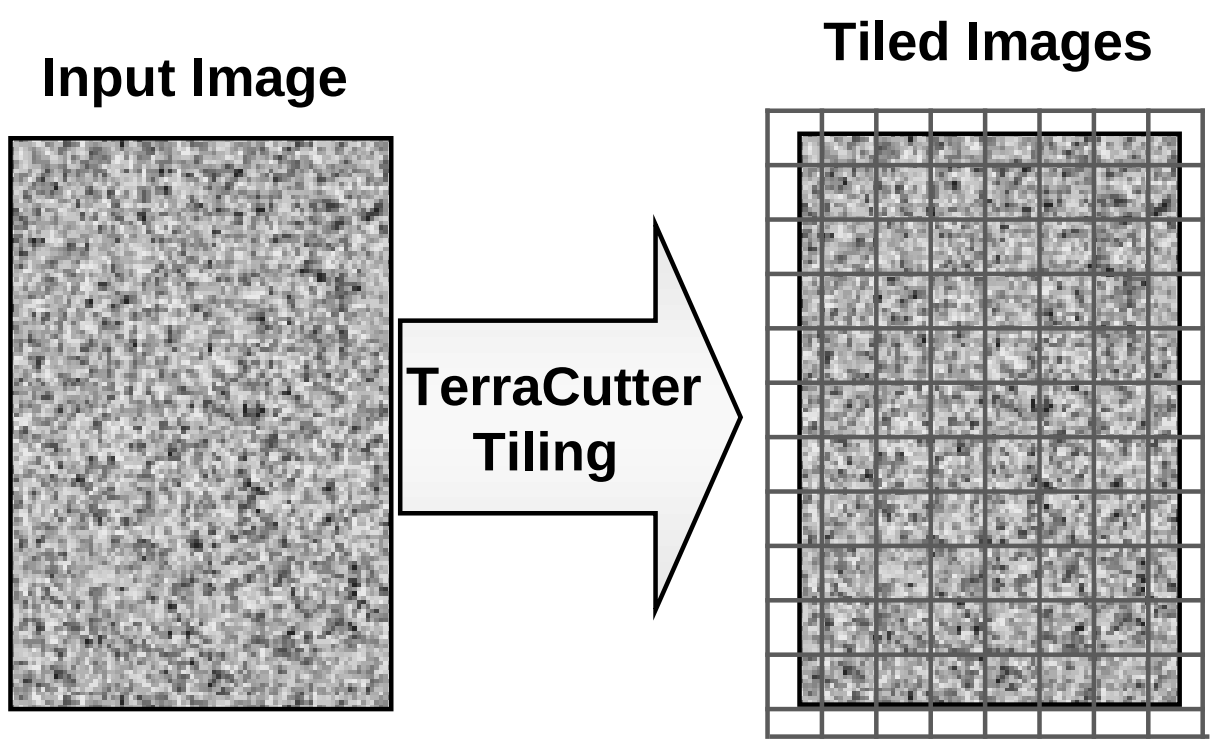

Figure 13: The TerraCutter breaks a large image up into tiles that then can be mosaiced into a scene.

### Raw Data-Set Tiling

For raw data-sets, TerraCutter begins cutting at logical pixel 0,0 in the top-left image file forming a single scene.[4] The input image is re-sampled to the appropriate resolution as it is read in. TerraCutter cuts the tiles and saves them in an uncompressed form (TIFF format) in a temporary directory tree. The input scene is padded with white space along the right and bottom edge so that all tiles are a fixed 200 x 200 pixels and no input pixels are lost.

The filename assigned to each tile identifies the tile's X, Y, and SceneID address. The X value begins at 0 and increments by one as TerraCutter tiles left to right. The Y value begins at the maximum number of tiles cut top to bottom and decrements by one as TerraCutter tiles top to bottom. The SceneID value is a unique integer assigned to the scene. It is a simple count of Scenes inserted into the TerraServer database.

Because a scene can be comprised of multiple overlapping input images, there is a possibility that a duplicate filename will be generated. When TerraCutter detects a duplicate filename, it merges the two tile images on a pixel by pixel basis. TerraCutter selects the pixel from the tile containing a non-white value. If both pixels are non-white, then the pixel from the input image currently being tiled is selected over the pixel from the duplicate file found on-disk. The resulting tile replaces the existing tile on-disk. It is possible that the process will be repeated again for another image in the scene.

Once an entire scene has been tiled, TerraCutter scans the temporary directory location and begins inserting tiles into the database. Meta-data for each tile is generated, and the TIFF image is compressed into a Jpeg or GIF file and inserted into the database using the ODBC APIs. Upon completion, TerraCutter marks the associated row in the *LoadJob* table as "completed". It inserts a new row in the *ScaleJob* table to schedule a pass over the inserted image tiles by the TerraScale program.

### Projected Data-Set Tiling

The database insertion process is more complicated for DOQ and DRG themes which mosaic many images together into one scene. TerraCutter must combine pixels from multiple input images into one tile. The merge must ensure geographic alignment so that roads, buildings and other structures that cross tile boundaries do not appear interrupted. TerraCutter does this by carefully computing the starting point - location 0,0 in the image tile. For UTM based data-sets (USGS DOQ and DRG), TerraCutter looks for the first pixel in the input image that has a UTM X and Y address that is evenly divisible by width and height of an image tile. For example, USGS DOQ images are 1-meter resolution, so DOQ tiles start at 200-meter boundaries. DRG images are 2-meter resolution, so DRG tiles start at 400-meter boundaries.

Rounding the starting UTM X and Y coordinate up to width and height of the image tile simplifies aligning layered maps containing multiple TerraServer data-sets. The UTM address for pixel 0,0 in a DOQ Image Tile at 2-meter resolution is the same UTM address for pixel 0,0 in a DRG tile with the same X, Y, zone address.

[4] Some data providers deliver a single theme as a set of image files. Others provide a scene in a single file. TerraCutter supports the ability to treat multiple images as one "virtual image".

Input image files of projected data-sets, like USGS DOQ and DRG, will overlap other image files along the edges. TerraCutter must choose which input image to take a duplicate pixel from. The amount of overlap varies from file to file in each data-set. Figure 13 depicts how input imagery files, numbered and outlined with solid thick lines, overlap each other within the UTM coordinate system. The tiles, outlined with light dashed lines within the numbered rectangles, depict the challenge in edge matching.

DOQ image files typically overlap each other by 100 to 300 pixels. DRG image files can overlap each other by 50 to 1500 pixels. However, only one file will contain "map data" while the others will contain map notes and tick marks found along the border of USGS topographical maps [Moore].

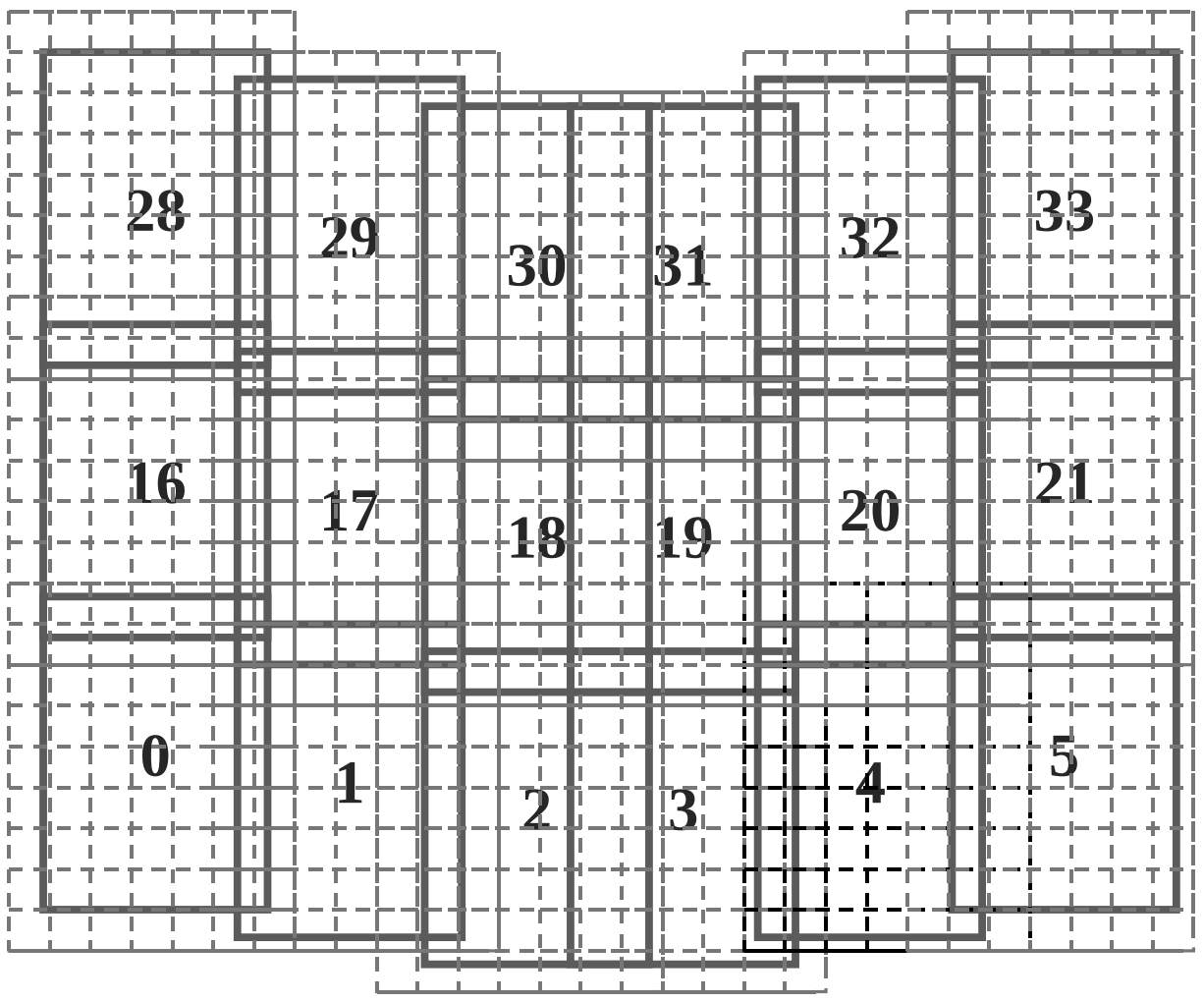

Figure 14. Tiling overlapping UTM images into a seamless mosaic. The dotted lines represent the mosaic grid. Up to four files can contribute pixels to a tile.

Conceptually, it is possible for up to four input images to contribute to a single TerraServer tile. To complete all the tiles for a single input image, a total of nine input images are needed – the center image and eight surrounding images. Unfortunately, the projected data-sets are not delivered in sorted order. Finding all the adjacent input images would be a tape shuffling nightmare. Hence, TerraCutter uses an incremental load algorithm.

TerraCutter tiles each input image independently. White space is added around the input image edge to align to the TerraServer grid system and the input data is re-sampled to the appropriate TerraServer resolution. Tiles are then cut and compressed to a temporary disk file.

After compressing each tile, TerraCutter looks for a tile with the matching Theme, Scale, X, Y, and SceneID properties in the appropriate TerraServer database imagery table. If there is not an existing tile, then TerraCutter inserts the image into the table and sets a "visibility flag" to "visible".

If a tile does exist in the database, TerraCutter compares the "blankness" of the newly cut tile with the tile in the database. If the new tile does not contain any white space from the input image edges, then the new tile is inserted, made visible, and the old image is set to "invisible". If the new tile does contain some amount of white space, but the tile in the database does not, TerraCutter discards the new tile and does not load it. If both tiles contain white space, TerraCutter fetches the old tile from the database, decompresses it, and does a pixel level merge with the old and new tile. The "blankness" of the resulting tile is computed, the merged tile is inserted into the database and made visible, and the old tile is marked invisible.

TerraCutter performs all four steps in one transaction – (1) check for an existing image, (2) merge pixels, (3) insert new tile row, and (4) update old tile's visibility flag. Using SQL Server concurrency control, other executing TerraCutters are automatically blocked from modifying the same tile, but can be updating other tiles in the same table. The TerraServer web application performs "dirty reads" of the imagery tables and is not blocked from reading the currently visible row. Thus, we are careful to change the visibility flag of the old tile as a last step so that the web application can get to a valid, but soon to be replaced tile, when TerraCutter is at step 2 or 3.

Once TerraCutter completes the tile insert, it deletes the temporary on-disk copy of the compressed tile. The program proceeds on to the next tile and repeats the process. When all tiles are cut from an input image file, TerraCutter updates the production status field in the Theme's *OriginalMeta* row to indicate that the input image has been completely tiled. TerraServer Administrators monitor the progress of the TerraCutter program through database queries against the Theme *OriginalMeta* table.

Should the TerraCutter program abort or be terminated before completion, the program will restart and pick up the tiling process from where it left off. The program uses the *ProdStatus* field in the Theme *OriginalMeta* table to determine if it finished an input image file. It skips through all the completed images until it finds the input image it was working on previously. It repeats the tiling process, but skips loading all tiles that were previously loaded.

## 5.4. TerraScale

TerraScale re-samples the tiles created by TerraCutter to create the lower resolution tiles in the theme's image pyramid. To create a lower resolution tile, TerraScale takes four tiles from the next higher resolution and averages four pixel values into one pixel value. TerraScale repeats this process at every resolution level until it tiles the lowest resolution tile for a given theme

Figure 15 depicts how the highest resolution tiles loaded by TerraCutter contribute to the pixels at lower resolution. We refer to the tiles loaded by the TerraCutter program as the "base scale" or "base tiles".

The number of lower resolution tile levels created by the TerraScale program is theme dependent. The USGS DOQ and SPIN-2 data base scale is 1-meter resolution. TerraScale creates 2- meter resolution through 64-meter resolution – a total of seven levels.

USGS DRG data is a special case. Original DRG input images are available at 2.5-meter resolution, 10-meter resolution, and 25-meter resolution. TerraCutter re-samples the 2.5-meter resolution image into the 2-meter tile table, the 10-meter input image into the 16-meter tile table and the 25-meter image into the 64-meter tile table. TerraScale re-samples the 2-meter tiles into the 4- and 8-meter tables. The 16-meter tiles are re-sampled into to the 32-meter table. The 64-meter tiles are re-sampled into the 128-meter table.

The TerraScale program recognizes base tile table changes. It handles the case where lower resolution tiles must be padded with blank space because not all of the higher resolution tiles yet exist. It determines that a lower resolution tile must be re-sampled based on the insert dates maintained in each tile table.

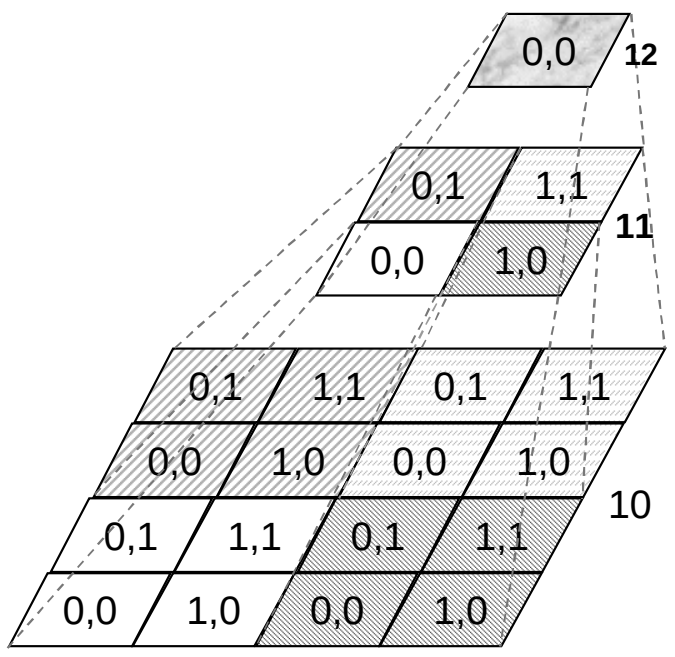

Figure 15: The TerraScale program computes the image pyramid by using averaging or using nearest neighbor algorithms on the four lower-level pixels.

The TerraScale program is designed so that multiple instances of the program can be running in parallel. Normally, one TerraScale program handles one zone of a projected theme or all scenes for a raw theme.

The TerraScale program continuously scans the *ScaleJob* table for new work to do. If it finds a row with a “job queued” status for the theme and zone it is assigned to, then it reads the job characteristics and updates the *ScaleJob* row to indicate that the TerraScale program is handling the job.

The *ScaleJob* fields identify the minimum and maximum X and Y values inserted into the base tile table by the TerraCutter for a single SceneID value. TerraScale computes the range of X and Y values for the lowest resolution scale that it must generate for the theme type. This constrains the size of the image pyramid the TerraScale program will attempt to create during this *ScaleJob* run.

TerraScale begins a loop to create the tiles at the lowest resolution. This is the top of the image pyramid. It fetches the current tile’s insert date and image pixels. Then it fetches the four tiles and insert dates at the next higher resolution level. It recurses down the image pyramid (higher resolution levels) until it fetches the four base tiles and the insert date. If any of the insert dates along the way are after the insert date found in the *ScaleJob* table, the TerraScale program will resample the imagery along the line of descent. If the dates of the higher resolution images are on or before the insert date found in the *ScaleJob*, the TerraScale can skip the re-sample process.

The TerraScale program continues to walk up-and-down the image pyramid underneath the lowest resolution tile it is generating. It finally completes and moves on to the next X,Y value to process for the *ScaleJob* and repeats the process.

TerraScale is told which resolution levels are to represent the image pyramid in the search system. As a last step in building an image pyramid for a particular low resolution X,Y value, TerraScale inserts the appropriate rows into the *Image* and *ImageSearch* tables. A tile is not visible in the TerraServer application until a row is inserted into these two tables.

When all the X, Y pairs are completed for the lowest resolution tiles, TerraScale updates the *ScaleJob* to indicate that it has completed the job. TerraServer System Administrators monitor the progress of TerraScale programs.

# 6. What We Learned

## 6.1. Initial Results

The TerraServer project began in late 1996. A prototype was demonstrated in May 1997. Aerial Images went live with a demonstration web site in January 1998. The full site went live in June 1998. It has now been operating for over a year.

When the web site was launched on June 24, 1998, it was overwhelmed with 35 million "hits". We had clearly under-estimated the popularity of this type of data.

Working with the hardware partners and the SQL Server development team, we configured the hardware and tuned the system software to handle 40 million hits and 300,000 visitors per day.

## 6.2. Traffic Analysis

To this day, TerraServer continues to be a very popular web site. Below are the usage statistics for TerraServer's first year on the web:

Table 2: TerraServer traffic summary July 1998 to July 1999.

| | **Total** | **Average/day** | **Max/day** |
|---|---|---|---|
| **Users** | 23,104,798 | 63,128 | 149,615 |
| **Sessions** | 31,011,284 | 84,730 | 172,545 |
| **Hits** | 2,287,259,402 | 6,624,607 | 29,265,400 |
| **Page views** | 367,528,901 | 1,004,177 | 6,626,921 |
| **DB queries** | 2,015,132,166 | 5,505,826 | 17,799,309 |
| **Image xfers** | 1,731,338,052 | 4,704,723 | 14,984,365 |

Since the launch, Microsoft TerraServer has reached a steady state of 5 to 8 million web hits, 5 to 6 million database stored procedure executions, and 50 GB of image tile downloads per day.

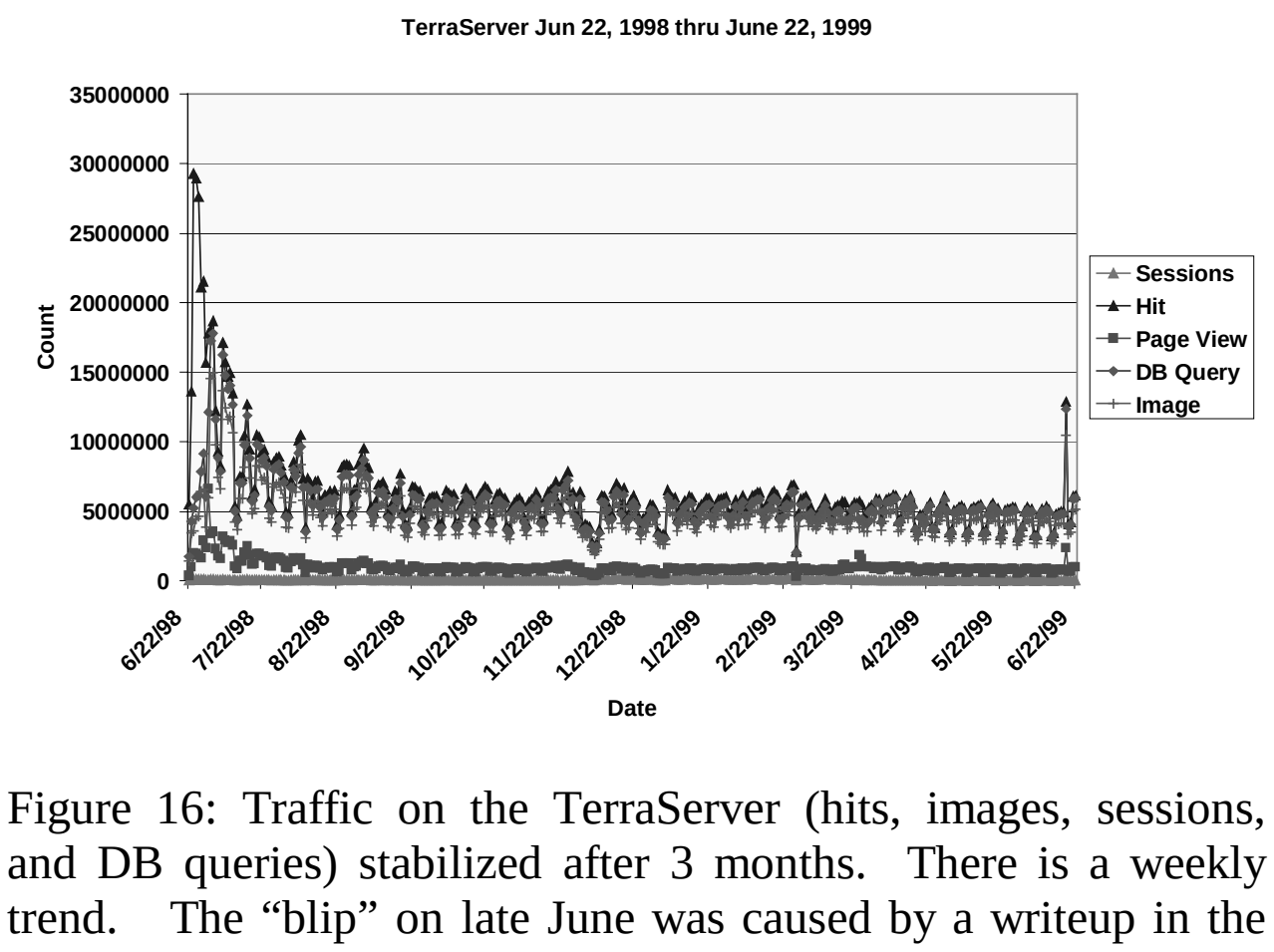

Figure 16: Traffic on the TerraServer (hits, images, sessions, and DB queries) stabilized after 3 months. There is a weekly trend. The "blip" on late June was caused by a writeup in the *Economist*. The drop in December appears to reflect vacations.

## 6.3. User Input

We received over 15,000 mail messages from users. We tried to answer each one. Most messages were constructive criticism or praise, but there were substantial complaints as well. The most common complaint was that images were missing. The server has only 30% coverage of the continental US and very spotty coverage outside the US. The second most common complaint was that images did not align. This forced us to go to the "scene" oriented design described here. The third most common complaint centered on the Java applet we wrote: it was difficult for us to get that applet to work on the many different Java Virtual Machines of the common platforms (each platform has many JVM variants).

## 6.4. System Availability

The Compaq database server and SQL Server 7.0 database management system have been extremely reliable. Table 3 summarizes the availability statistics for the period through 1 July 1998 through 1 July 1999. The system went out of service for 3 hours for software upgrades, 2.5 hours to move the server within the data center, and 33 minutes due to a software bug.

Table 3: Availability statistics for Microsoft TerraServer SQL Server.

| | Hours | % |
|---|---|---|
| Elapsed Time | 8760 hours | 100.00% |
| Availability | 8754 hours | 99.93% |
| Scheduled Availability | 8757 hours | 99.97% |

## 6.4. Database Size and Performance

Table 4 summarizes the database size as of July 1999. The current configuration has the capacity to double the database size.

Table 4: Database size.

| Item | Count | Rows | Total Size |
|---|---|---|---|
| Database Files | 81 | - | 1,600 GB |
| Log File | 1 | - | 18 GB |
| Image Tables | 18 | 134 million | 1365 GB |
| Gazetteer Tables | 8 | 109 million | 3.3 GB |
| Load Mgmt Tables | 4 | 72,000 | 25 MB |

The 1.1 TB database is backed up regularly to the StorageTek 9710 TimberWolf tape robot using SQL Backup integrated with Legato Networker. In on-line mode, the backup consumes approximately 20% of the CPU resources and takes approximately 11 hours to complete including tape changes. In off-line mode, Legato Networker can backup the entire 1.1 TB database in 7 hours, 15 minutes. The backup fills 32 DLT 7000 tapes.

### 6.5. Application Size and Complexity

Table 6 summarizes the size of the TerraServer application source code:

Table 6: Web Active Server Pages and SQL Server Stored Procedures.

| Item | Modules | Files | Lines |
|---|---|---|---|
| T-SQL Stored Procedures | 47 | 47 | 7677 |
| Active Server Pages (Tier 2 Web App) | 32 | 49 | 6100 |
| Load Programs | 2 | 73 | 34123 |

One full-time developer and 4 part-time developers built TerraServer. 1 full-time system administrator and 1 full-time data-load specialist maintain the web site.

### 6.6. New Satellite/Aerial Imagery Users

Remote sensing and aerial photography have been a niche application due to the high complexity and expense of tools that can view it. Microsoft TerraServer dramatically reduced the access complexity and cost for simple applications. The application is so simple to use that, high-resolution imagery is now available to the entire Internet population. We and others have been astonished at the wide interest in the data: hundreds of thousands of people visit the site each day (see Table 2).

### 6.7. The Internet as a Data Distribution Mechanism

Historically, high-resolution imagery was distributed on magnetic tape or CDROM. The time from deciding what imagery was needed to getting the data was measured in weeks. With TerraServer, users can see exactly what they are getting. The purchased area can be specified more accurately. And with the Internet, the data can be transferred electronically, eliminating physical media transfer. The turn around time is reduced from weeks to minutes.

This convenient access is a key part of the site's popularity.

### 6.8. Relational Databases as Image Repositories

Using relational databases to store image pyramids of common graphics file formats, e.g. Jpeg and GIF, forces the separation of storage management from image presentation. The tiled design allows rapid pan and zoom to any part of the image database. It also supports background loading of new images while the current data is being viewed. The database system is able to handle much larger image bases than a file-per-image design used by earlier efforts. Storing tiles individually also allows for easy on-line editing of any portion of an image. Choosing a ubiquitous medium like the Internet and the common web browser as the presentation tool enabled the rapid dissemination of high-resolution imagery to new users and applications.

### 6.9. The Value of Cooperative Joint Research

Because the project had to use real data, and that data was expensive, it forced us to enlarge the project team beyond database and systems researchers. By including additional companies and organizations, the project goals and requirements expanded. This brought additional skills to the table – geographers, graphics researchers, high-resolution image interpreters (a.k.a. spies), and GIS experts. We were able to blend the knowledge and skills of diverse partners to build a powerful spatial data warehouse and produce a more complete result by solving a wider set of problems than just a database or operating system problem.

### 6.6. Integration With Encarta Online

TerraServer became part of the EncartaOnline web site in May 1999. The Microsoft Encarta product team cross-referenced Encarta Encyclopedia articles with TerraServer imagery. As users navigate the imagery, hyper-text links appear to related Encarta Encyclopedia articles.

This vastly improves the richness of the user interface.

# 7. Future Work

## 7.1. Layered Maps

We are collaborating with UC Berkeley Digital Library Project, http://elib.cs.berkeley.edu/, on layered maps. The USGS DOQ and DRG data-sets are in a common projection system. The TerraServer tiling algorithm cuts tiles so that client applications can identify overlapping tiles from separate themes. We plan to work with the UCB Digital Library team to build a client application which will display TerraServer projected data-sets that are in the same projection as a layered map set. The layered map user interfaces has the same ease-of-use goals as our traditional single-layer HTML interface

## 7.2. More Themes

The popularity of the web site has encouraged other data providers to offer interesting data sets. We plan to add natural color aerial imagery of Western Europe and Great Britain. Some of these data-sets will be ½ meter resolution data.

## 7.3. Temporal Data

The TerraServer lets uses move through space, but next they will want to watch a movie of a particular area as it evolves over time.